\documentclass[preprint,showpacs,preprintnumbers,amsmath,amssymb,aps,prb,floatfix]{revtex4-1}

\usepackage{graphicx}
\usepackage{dcolumn}
\usepackage{bm}

\input colordvi

\begin{document}


\preprint{DRAFT --- LA-UR-13-21965}

\title{Density Functional Theory Study of 
the Structural Properties of  PuH$_x$, $2\le x\le 3$}

\author{Sven P. Rudin}
\affiliation{Los Alamos National Laboratory, Los Alamos, New Mexico 87545, USA}

\date{\today}

\begin{abstract}
Plutonium dihydride and trihydride show strikingly similar crystal structures
when viewed as close-packed Pu planes with ABC and AB stacking, respectively.
The similarity serves as a framework for density functional theory (DFT)
calculations of PuH$_2$, PuH$_3$, and intermediate compositions.
Agreement between structures observed in experiment and
in the DFT description of the Pu-H system
requires accounting for the strong electronic correlation in the $f$ orbitals,
achieved here with the addition of a Hubbard parameter $U$.
The hysteresis measured between hydriding and dehydriding
can be attributed to the effect of stacking of the close-packed Pu planes on
the energy as a function of stoichiometry, calculated using the GGA+U approach.
Changes in the interstitial positions occupied by the H atoms affect
the energy by amounts that are negligible compared to room temperature;
changes in the magnetic structure lead to equally small
energy differences.
\end{abstract}

\maketitle

\section{Introduction}

The safe storage of plutonium will require a better understanding of the
metal's corrosion.
Hydrogen-catalyzed corrosion is of greatest concern,\cite{martz03}
given that plutonium hydride in particular not only forms easily,\cite{richmond2010}
but is also highly reactive.\cite{haschke91, haschke2001}
Yet  the solid Pu-H system has not received much theoretical study,
and such studies have focused on the two stoichiometric
compounds,\cite{eriksson91,Ai2012, Ao2012jnm}
with limited attempts to connect them with intermediate compositions.\cite{Ao2012cpl}
Experimentally much more has been done.\cite{mulford55, mulford56, aldred1979,
ward1983, willis1985, haschke1987, mcgillivray2003, richmond2010}

Known Pu-H system stoichiometries range from the pure metal
through the dihydride to the trihydride.
Hydrogen shows solubility in Pu for up to 1-2 at.\% H.\cite{richmond2010}
Adding additional H into the system leads to a region wherein the Pu metal and
the dihydride coexist,
the upper limit is a temperature-dependent
PuH$_x$ stoichiometry with $x=1.75-1.88$.\cite{mulford55}
Here a second region of H solubility appears, now in PuH$_2$.
The upper limit of this region, PuH$_{2.75}$, marks the beginning of
a region of coexistence between the
dihydride and the trihydride.\cite{mulford56}

This phase transition connecting di- and trihydride shows a large hysteresis effect
between cycles of hydriding and dehydriding.\cite{mulford56}
The transformation proceeds more easily for dehydriding than for hydriding
with the largest difference in the H pressure (a factor of $10^4$)
near $x=2.9$.\cite{haschke1987}
Experimentally, the hysteresis appears to have structural origins.

The current work explores the structures of the di- and trihydride
in terms of how they relate to each other and
how they change for intermediate stoichiometries.
Section \ref{sec:crystal} reveals remarkable similarities between
the di- and trihydride crystal structures.
The ensuing sections use the similarities to construct a
framework and describe calculations aimed
at understanding the hysteresis.

\section{Crystal Structure Similarities \label{sec:crystal}}

\begin{figure} 
\includegraphics[width=8.5cm]{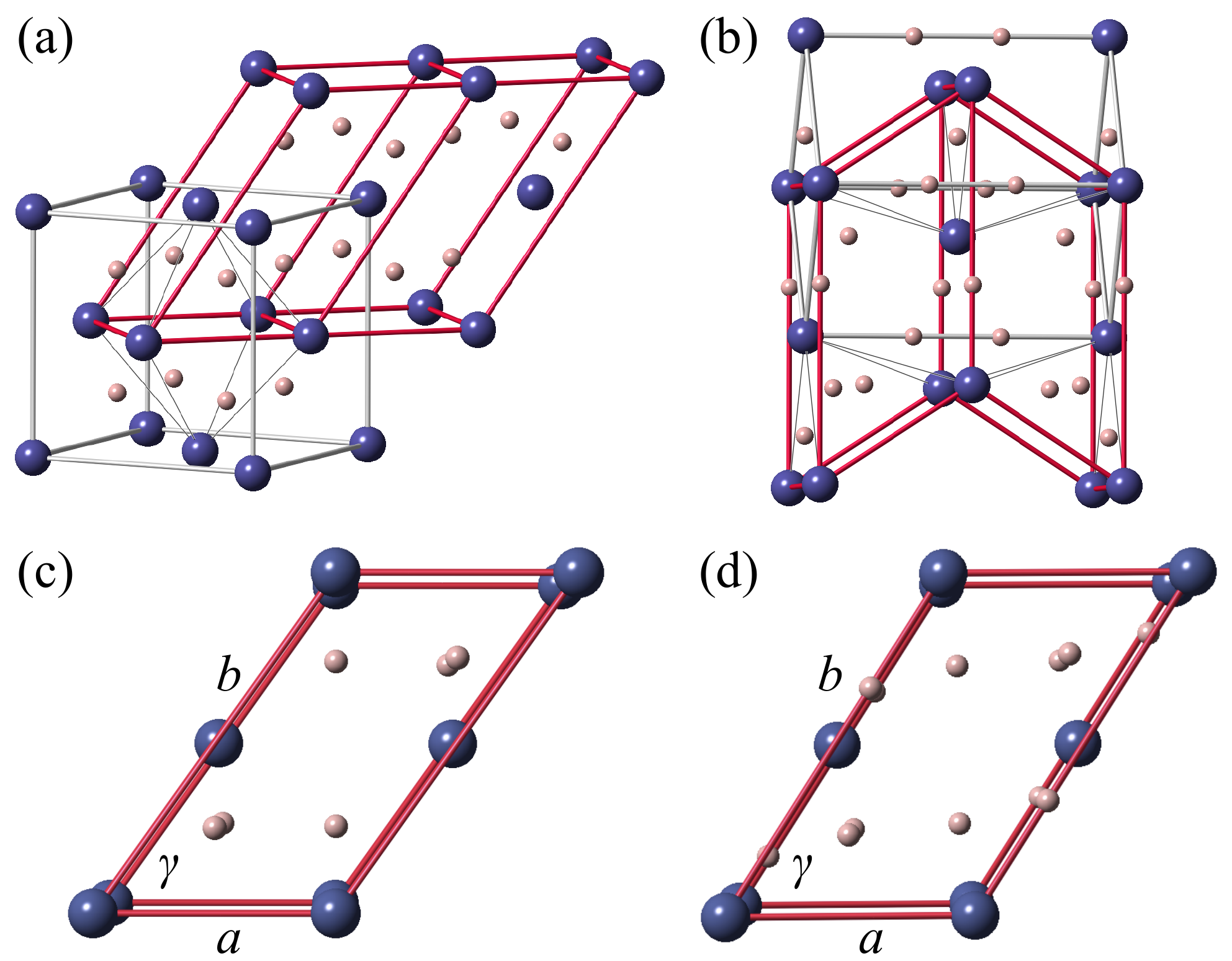}
\caption{\label{fig:relac}
(Color online)
Monoclinic structural units (dark red edges) of (a) PuH$_2$ and (b) PuH$_3$
as they relate to the conventional unit cells (grey edges).
Large spheres represent Pu atom sites, small spheres represent H atom sites;
edges of the octahedra formed by Pu sites appear as thin lines.
The lattice of PuH$_2$ repeats a monoclinic structural unit (MSU),
while that of PuH$_3$ repeats a MSU and its mirror image.
The top view (with $c$ edge into the page) of the MSUs
for (c) PuH$_2$ and (d) PuH$_3$
emphasizes the similarity between the two systems:
identical Pu positions within the MSUs, and H positions within the MSUs
that differ almost only in the
additional H atoms in the faces spanned by $b$ and $c$ for PuH$_3$
(the close-packed planes).
}
\end{figure}

While the di- and trihydride crystal structures have very different symmetries,
their Pu sublattices can both be viewed as close-packed lattices.
The cubic
PuH$_2$ appears in the fluorite structure (Fig.~\ref{fig:relac}(a));
the hexagonal
PuH$_3$ is isostructural to Na$_3$As (Fig.~\ref{fig:relac}(b)).
The Pu sublattice of PuH$_2$ corresponds to a face-centered cubic (fcc)
crystal structure,
making it close-packed layers with ABC stacking.
The Pu sublattice of PuH$_3$ takes on a hexagonal close-packed
(hcp) crystal structure,
making it close-packed layers with AB stacking.
These structures echo the layered, close-packed nature already
apparent in crystal structures of pure Pu:
the face-centered cubic structure of the Pu sublattice of PuH$_2$
is that of $\delta$-Pu,
and the AB stacking of Pu sublattice of PuH$_3$ can be related to
the $\alpha$-Pu structure's
repeating two planes of a distorted hexagonal structure.\cite{crocker1971}

The di- and trihydride crystal structures have very different primitive unit cells
but can also be constructed from structural units with
surprisingly comparable dimensions and internal structure.
These are monoclinic structural units (MSUs).
Figure \ref{fig:relac} shows a description of
(a) the PuH$_2$ lattice in terms of repeated MSUs and
(b) the PuH$_3$ lattice built from a similar MSU and its exact mirror image.
Each MSU shifts the Pu positions from one plane to the next,
in ABC stacking the shift simply repeats,
whereas in AB stacking the direction of the shift alternates.

Table \ref{tab:msudim} shows the close proximity of the experimental dimensions of the
monoclinic structural units for PuH$_2$ and PuH$_3$.
The experimental Pu-Pu distances within the planes are nearly identical for PuH$_2$ and PuH$_3$,
as evidenced by the in-plane lattice vectors $b$ and $c$, which differ between the
structures by only 0.01~\AA.
The angle $\gamma$ also differs by only two degrees between structures.
The single large difference appears for $a$: the spacing between layers is
6\% larger for PuH$_3$ to accommodate the additional H atoms.

\begin{table}
\begin{center}
\begin{tabular}{c c | c | c c c c | c }
\multispan2{\hfil crystal\hfil}  & method  &  $a$ & $b$ & $c$  & $\gamma$    & $V$     \cr
\hline
\hline
      PuH$_2$ & ABC & experimental & 3.79 & 6.56 & 3.79 & 55$^\circ$  & 77.0  \cr
      PuH$_2$ & ABC & GGA, AFM &  3.78 & 6.57 & 3.72 & 54$^\circ$ & 74.5 \cr
      PuH$_2$ & ABC & GGA with SOC, AFM & 3.77 & 6.54 & 3.72 & 54$^\circ$ & 74.2 \cr
      PuH$_2$ & ABC & GGA+U, AFM &  3.84 & 6.67 & 3.82 & 54$^\circ$ & 79.5 \cr
\hline      
      PuH$_2$ & AB  & GGA, AFM &  3.38 & 6.90 & 3.34 & 60$^\circ$ & 67.5 \cr
      PuH$_2$ & AB  & GGA with SOC, AFM & 3.43 & 6.87 & 3.31 & 60$^\circ$ & 67.4 \cr
      PuH$_2$ & AB  & GGA+U, FM & 3.61 & 6.95 & 3.77 & 56$^\circ$ & 78.5 \cr
\hline
\hline
      PuH$_3$ & AB & experimental &  4.02 & 6.55 & 3.78 & 57$^\circ$ & 83.6 \cr
      PuH$_3$ & AB & GGA, FM &  3.96 & 6.45 & 3.72 & 57$^\circ$ & 79.9  \cr
      PuH$_3$ & AB & GGA with SOC, FM & 3.95 & 6.45 & 3.71 & 57$^\circ$ & 79.3 \cr
      PuH$_3$ & AB & GGA+U, AFM & 4.07 & 6.61 & 3.82 & 57$^\circ$ & 86.2 \cr
\hline      
      PuH$_3$ & ABC  & GGA, FM &  3.70 & 6.41 & 3.70 & 55$^\circ$ & 71.6 \cr
      PuH$_3$ & ABC & GGA with SOC, FM & 3.68 & 6.44 & 3.69 & 55$^\circ$ & 71.7 \cr
      PuH$_3$ & ABC  & GGA+U, AFM &  3.71 & 6.80 & 3.70 & 56$^\circ$ & 77.5 \cr
\end{tabular}
\caption[]{Measured and calculated dimensions of monoclinic structural units
in PuH$_2$ and PuH$_3$.
Lattice parameters $a$, $b$, and $c$ are given in \AA, volume $V$ in \AA$^3$.
The angle $\gamma$ describes the slant of the MSUs (see Fig.\ \ref{fig:relac})
and is not related to the hexagonal nature found in the close-packed Pu layers.
Results are reported for magnetic structure with the lowest energy for the
particular stoichiometry, lattice structure, and method.
All GGA+U calculations employ $U=4$ eV.
}
\label{tab:msudim}
\end{center}
\end{table}

The internal structure of the MSUs for PuH$_2$ and PuH$_3$
also appear remarkably similar.
The Pu sites are identical.
The H atoms in PuH$_2$ sit at the centers of tetrahedra formed by Pu atoms.
These tetrahedral interstices appear in planes located at $\frac{1}{4} b$ and $\frac{3}{4} b$
in the MSU of Fig.~\ref{fig:relac}(c).
Corresponding H atoms appear on almost unchanged sites
within the MSU of PuH$_3$ (on the 4f Wyckoff positions)
and are joined by H atoms on the faces spanned by $b$ and $c$
(2b positions).
Within the primitive unit cell of PuH$_3$ these two types of H sites also have different
environments,
the former still sit in Pu tetrahedra, albeit slightly deformed,
the latter sit at the centers of equilateral Pu triangles.

\section{Calculation Method \label{sec:method}}

The similar MSUs of Pu di- and trihydride provide
a framework within which intermediate compositions can be explored computationally.
The calculations presented here are limited to unit cells with four Pu sites
and between eight and twelve H atoms.
Initial unit cells correspond to the two MSUs shown in Fig.~\ref{fig:relac}(a)
and in Fig.~\ref{fig:relac}(b) (where the chevron shape is made into an actual unit cell)
with added and removed H atoms, respectively.
These initial configurations often relax surprisingly slow with false plateaus,
which require rather small convergence criteria to overcome (see below).
These unit cells limit the current study to structures with AB and ABC stacking.
A systematic study of whether the Pu hydrides favor other stacking sequences
will likely shed more light on the system but also require significantly more
computational effort.

The calculations discussed here neglect thermal effects.
Contributions to the free energy from the phonons in particular are important,
given the small mass of the H atoms.
The current results do reveal many small energy differences between magnetic states
and various configurations of the H atoms, which even the zero-point energy
will affect.
However, in addition to being beyond the scope of this work, the calculation of phonons
and including their thermal effects will not likely overcome the inaccuracy inherent
in the density functional theory (DFT) method.
Connections made between the results and experimental data do involve
thermal effects in the form of comparisons with the energy scale set by room temperature,
but the main conclusions involve large enough energy differences to be immune
to neglected contributions from the phonons.

The results presented here originate in
DFT calculations using
the {\sc VASP} package.\cite{kresse96, kresse99}
The calculations make use of
the generalized gradient approximation (GGA) of
Perdew, Burke, and Ernzerhof.\cite{PBE96}
The Pu($5f, 6d, 7s$) and H($1s$) electrons are treated in the valence
with projector-augmented wave potentials.\cite{blochl94a}
The calculations employ the linear tetrahedron method
with Bl\"ochl corrections,\cite{blochl94}
a {\bf k}-point mesh of density 60~\AA$^{-1}$,
and an energy cutoff of  500 eV.
The self-consistent cycles are converged to within 10$^{-5}$ meV
to enable the 10$^{-4}$ meV ionic stopping criteria needed to overcome
the false plateaus mentioned above.
The calculations allow spin polarization and compare two types of magnetic structure,
ferromagnetic (FM) and antiferromagnetic (AFM).

Because DFT in the GGA 
fails to describe PuH$_3$ as a semiconductor,\cite{Ai2012}
the effects of including either strong electron correlation and spin-orbit coupling
are investigated.
Some results reported here stem from calculations that
treat the on-site Coulomb repulsion between 5$f$ electrons 
with a Hubbard parameter $U$ (GGA+U)
in the rotationally invariant form of Dudarev {\it et al.}\cite{dudarev98}
In this form the Hubbard parameter $U$ and the exchange parameter $J$ 
appear only in the difference $U-J$,
throughout this report the difference is simply referred to as $U$.
A single value for $U$ (4 eV) for all stoichiometries
allows a comparison of the calculated energies and the evaluation
of formation energies.
While Ai {\it et al.} suggest spin-orbit coupling (SOC) can be neglected,\cite{Ai2012}
the results presented here show some energy differences small enough
for SOC to matter.

\section{Results for PuH$_2$ and PuH$_3$}

\subsection{DFT in the GGA}

\begin{figure} 
\includegraphics[width=8.5cm]{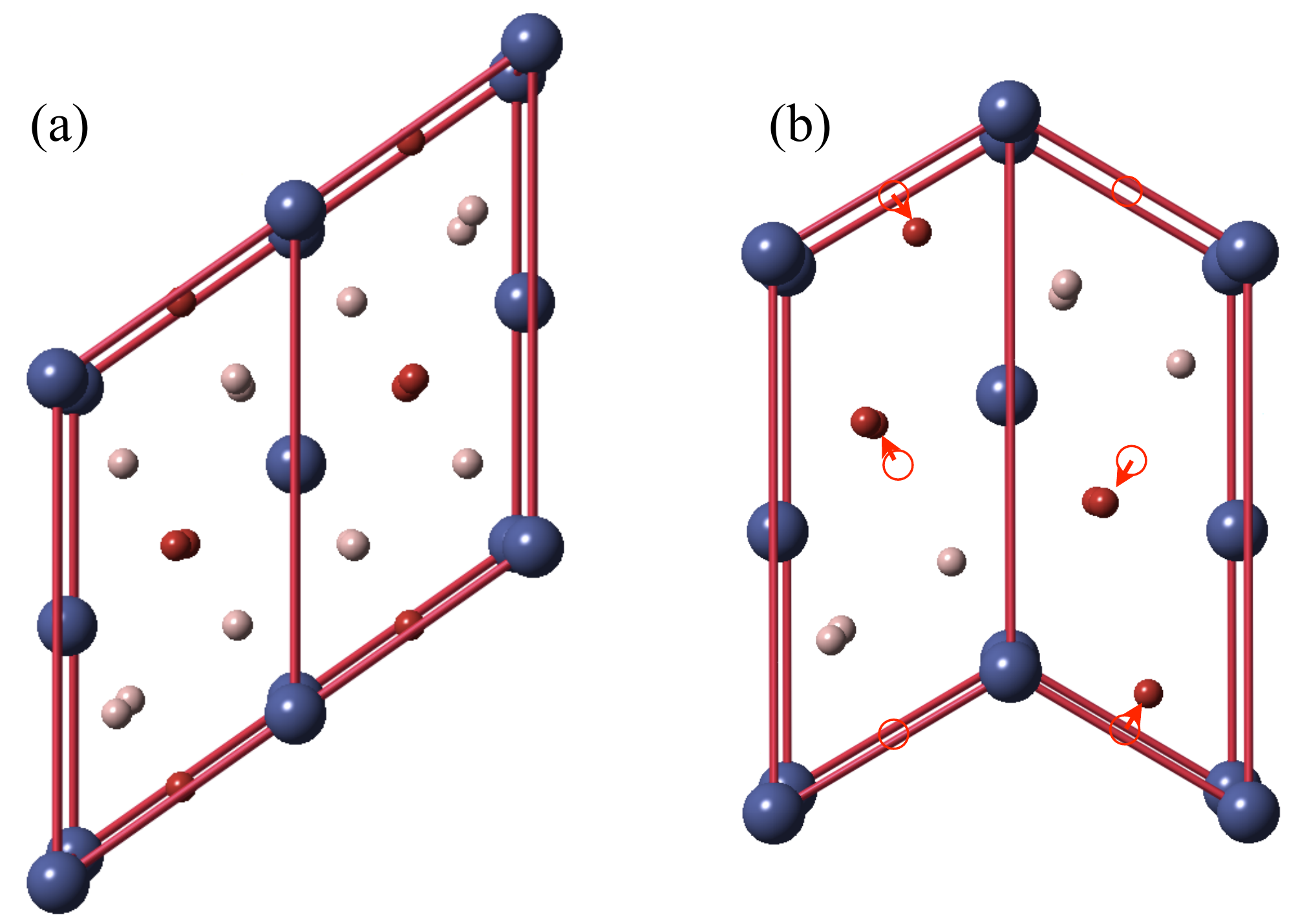}
\caption{\label{fig:alts}
(Color online)
Crystal structures with lowest energy in the GGA to DFT for
the stoichiometric Pu hydrides.
(a) The favored PuH$_3$ crystal structure
compares directly to the experimental crystal structure of PuH$_2$
(Fig.~\ref{fig:relac}(a) and (c))
with additional H atoms occupying octahedral sites (small dark spheres).
(b) The favored PuH$_2$ crystal structure
exhibits a Pu sublattice comparable to that of
the experimental PuH$_3$ crystal structure
(Fig.~\ref{fig:relac}(b) and (d))
with half of the H atoms sitting on the tetrahedral sites (small light spheres)
and the other half (small dark spheres) sitting near the octahedral sites
(empty circles).
}
\end{figure}

For both stoichiometric compounds,
DFT in the GGA without SOC fails to deliver the correct ground state crystal structure.
Curiously, the GGA would have the two hydrides switch the stacking
of Pu sublattices (see Fig.~\ref{fig:alts}):
DFT in the GGA would have PuH$_3$ take on a cubic symmetry, i.e., ABC stacking
of the Pu sublattice with H atoms at both the tetragonal and the octahedral sites;
an energy difference of 110 meV per formula unit (meV/f.u.)
appears relative to the experimental structure (both structures favor FM).
Similarly, DFT in the GGA would have PuH$_2$ with an AB stacking
of the Pu sublattice with H atoms sitting in the slightly deformed Pu tetrahedra
(the Pu sublattice does not quite have hexagonal symmetry);
an energy difference of 8 meV/f.u.\ appears relative to the experimental structure
(both structures favor AFM).
The value 8 meV/f.u.\ presents an example of an energy difference that
thermal energy stemming from the phonons could strongly affect.

\subsection{including spin-orbit coupling}

DFT in the GGA with SOC shows the correct ground state crystal structure
for PuH$_2$ but still fails for PuH$_3$.
The correct ABC stacking for PuH$_2$ lies 41 meV/f.u.\ lower than AB stacking
and favors AFM over FM by 61 meV/f.u.
PuH$_2$ remains a metal.
For PuH$_3$, the erroneous ABC stacking lies 75 meV/f.u.\ lower than the
correct AB stacking, and FM remains favored over AFM by 62 meV/f.u.

\subsection{including a Hubbard U}

Accounting for the electronic correlation in the $f$ orbitals describes
the structures of the stoichiometric compounds in agreement with experiment.
As shown by Ai {\it et al.},\cite{Ai2012} inclusion of the Hubbard $U$
in the calculations leaves
PuH$_2$ metallic with $f$ states at and below $E_F$,
while in PuH$_3$ the Hubbard $U$ opens a gap between occupied
and unoccupied $f$ states.
This approach (performed here without SOC)
also results in the correct structures for both stoichiometric compounds:
for PuH$_2$, GGA+U calculations favor the correct ABC stacking
over the erroneous AB stacking by 360 meV/f.u.;
for PuH$_3$, GGA+U calculations favor the correct AB stacking
over the erroneous ABC stacking by 140 meV/f.u.

In both stoichiometries,
the GGA+U calculations give lower energies for AFM than FM order.
The energy differences are quite small,
60 meV/f.u.\ for PuH$_2$ and 20 meV/f.u.\ for PuH$_3$.
These small values suggest a delicate
balance between magnetic structures
(which could easily be upset by either thermal effects or
other theoretical methods).
FM order appears in most experimental measurements
PuH$_{2\pm x}$ and PuH$_3$,\cite{ward1983, willis1985}
though earlier measurements suggest
AFM order for PuH$_2$.\cite{aldred1979}

\section{Results for PuH$_x$ with $2<x<3$}

Expanding on the results for the di- and trihydride,
the methods described above are applied to structures with AB and ABC stacking for
PuH$_x$ with $2<x<3$.
Systems with intermediate stoichiometries originate in
PuH$_2$ and PuH$_3$ unit cells (with four Pu atoms)
with added and removed H atoms, respectively.
Structural optimization of these initial structures proceeds until
the total energy between optimization steps differs by less than $10^{-4}$ meV;
this process frequently involves unusually numerous optimization steps.
Seeding with all reasonably different H placements
leads to multiple (meta)stable arrangements for each stoichiometry.

\begin{figure} 
\includegraphics[width=8.5cm]{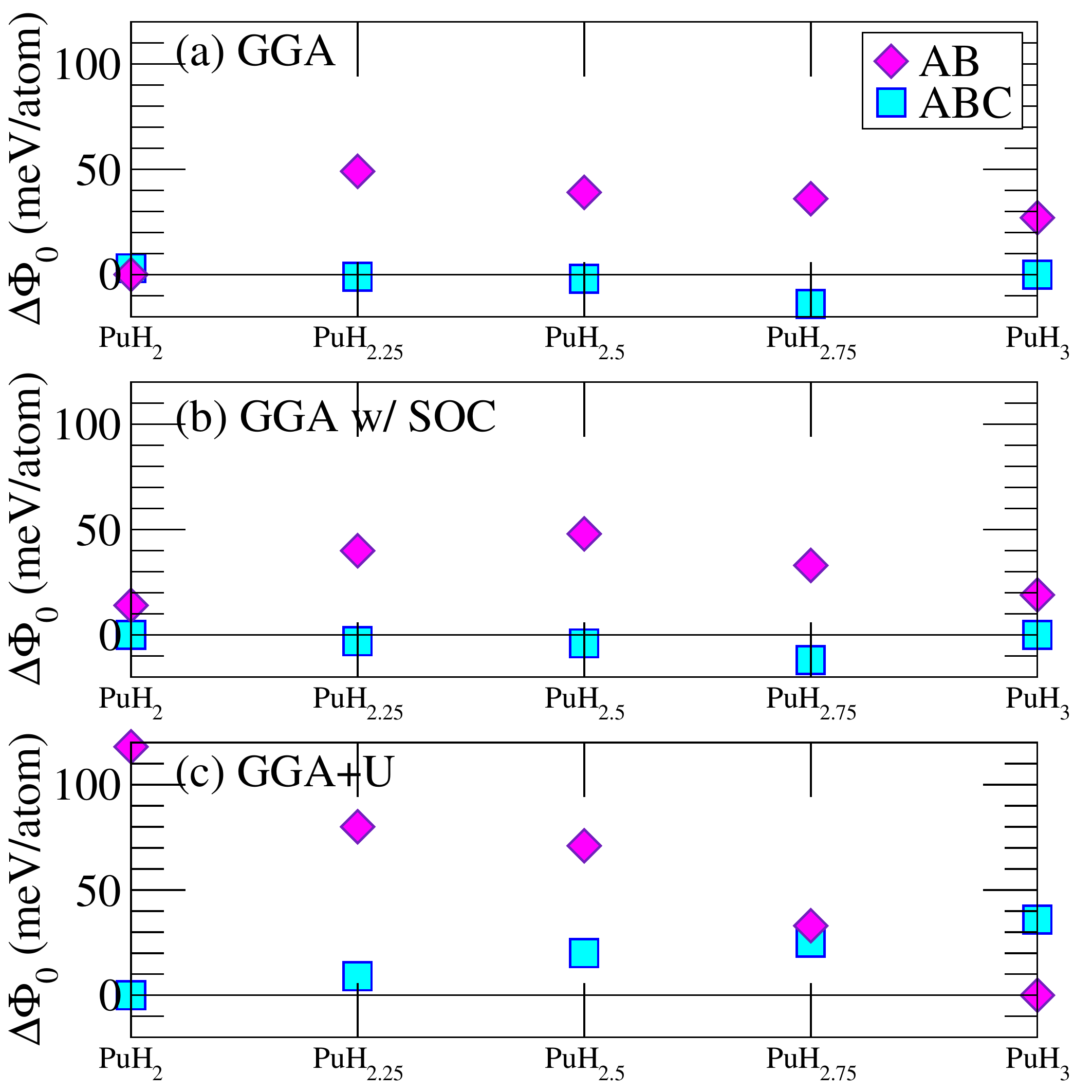}
\caption{\label{fig:FormEgga}
(color online)
Relative calculated potential energies
(with respect to PuH$_2$ and PuH$_3$) of
the most stable PuH$_x$ supercells
with ABC stacking (squares)
and AB stacking (diamonds).
Energies are given relative to the most favorable crystal structure
and magnetic structure of the di- and trihydride within each method:
(a) AFM PuH$_2$ with AB stacking and FM  PuH$_3$ with ABC stacking for GGA,
(b) AFM PuH$_2$ with ABC stacking and FM  PuH$_3$ with ABC stacking for GGA with SOC,
(c) AFM PuH$_2$ with ABC stacking and AFM  PuH$_3$ with AB stacking for GGA+U.
}
\end{figure}

Figure \ref{fig:FormEgga} shows the calculated formation energies
(relative to the stoichiometric compounds)
of PuH$_x$ structures with $2<x<3$.
The formation energies for PuH$_x$
structures relative to PuH$_2$ and PuH$_3$
derive from comparing to the linearly-weighted
energies of the di- and trihydride (i.e., to
$E(\text{PuH$_2$})+(x-2)[E(\text{PuH$_3$}) - E(\text{PuH$_2$})]$).
The energies at these endpoints are for the most favored crystal
and magnetic structure of the method in question,
as outlined in the figure caption.

Structures with ABC stacking have lower energies than those with AB
stacking, independent of method, with two exceptions.
(1) DFT in the GGA (without SOC) fails to reproduce the correct
crystal structure for PuH$_2$;
(2) in the GGA+U approach, DFT reproduces the
correct crystal structure for PuH$_3$.
This second exception makes only the GGA+U results relevant
for comparisons with experiment;
consequently the remainder of this section discusses only GGA+U results.
Interpolating the GGA+U results between the considered stoichiometries
suggests a region of stability for ABC stacking between PuH$_2$ and
just above PuH$_{2.75}$, followed by 
a region of stability for AB stacking starting slightly beyond PuH$_{2.75}$.

The GGA+U calculated formation energies from PuH$_2$ to PuH$_{2.75}$ remain below
room temperature without changing crystal structure.
This calculated thermally-accessible region fits the
experimentally observed region of H solubility in PuH$_2$,
for which PuH$_{2.75}$ demarcates the upper limit.
While these calculations reveal nothing about the kinetics of moving hydrogen into
the PuH$_2$ crystal,
they do suggest that, at room temperature, thermal effects will not push the hydrogen
back out.

The location of H atoms in the PuH$_{x}$ lattice suggested by
the current calculations agrees with experiment.
The calculated lowest energy for PuH$_{2.25}$, PuH$_{2.5}$, and
PuH$_{2.75}$ appears for structures with the additional one, two, or three
H atom(s) in octahedral interstitials.
Occupying additional octahedral interstitial sites (by creating
tetrahedral vacancies) raises the calculated energies by between 5
and 25 meV/atom.
This range of energy is easily attained at room temperature.
Finite temperature also affects the free energy by way of 
configurational entropy, which will favor partial occupancy of
both tetrahedral and octahedral sites.
Tetrahedral vacancies in concert with partial
occupancy of octahedral sites for PuH$_x$
have been observed experimentally.\cite{haschke1987}
This partial occupancy for both types of sites is also observed for
PuD$_x$,\cite{bartscher1984, bartscher1985}
for which the calculated results presented here hold equally well.

The magnetic structures found to be most favorable by the current calculations
do not agree with experiment, but energy differences are small.
The GGA+U calculations prefer an AFM structure for all stoichiometries.
Magnetic susceptibility measurements show FM behavior
for a wide range of stoichiometries,\cite{willis1985}
though earlier measurements suggest AFM ordering for the dihydride.\cite{aldred1979}
The calculated energy differences between AFM and FM ordering, however,
range from 5 meV/atom for PuH$_3$ to 15 meV/atom for PuH$_2$
(i.e., again an energy range easily attained at room temperature).

\section{Relating the GGA+U Results to the Hysteresis}

The small energy difference between AB and ABC stacking for PuH$_{2.75}$
in the GGA+U calculations warrants considering transformations between
the two crystal structures near this stoichiometry.
Exploratory calculations, mentioned here in passing, take the unit cell with AB stacking,
shear it stepwise and relax internal coordinates until the stacking is ABC;
these calculations suggest an energy barrier for this transformation
comparable to room temperature.
An accurate evaluation of the barrier height would require a nudged elastic
band calculation that optimizes internal and unit cell degrees of freedom.\cite{Sheppard2012}
This evaluation would have to be repeated for other pathways, including those
in other systems with, for example, larger unit cells to accommodate alternate
stacking sequences.
Such calculations would go beyond the scope of this work, and
furthermore would not represent macroscopic processes:
in a solid of measurable size, such a concerted mechanism
cannot be thermally activated.

The transformation more likely involves a locally nucleated process followed by
energetically-favored growth of the transformed region,
a conversion process which the current GGA+U results support.
The GGA+U calculated energy differences between AB and ABC stacking
at stoichiometries
away from PuH$_{2.75}$ can encourage growth of a
region of one type of stacking nucleated within the other type.
While the values of the necessary energy differences are unknown,
they are finite, which introduces a difference in the stoichiometries that
foster the two transformations.
This difference in stoichiometries which drive
AB$\rightarrow$ABC and
ABC$\rightarrow$AB transformations
is the source of the hysteresis measured between hydriding and dehydriding.

\section{Summary}

Plutonium dihydride and trihydride show strikingly similar crystal structure aspects
when viewed as close-packed Pu planes with ABC and AB stacking, respectively,
with interstitial H atoms.
The dimensions within the planes differ by only 0.01 \AA,
the planes are spaced 6\% farther apart in PuH$_3$ compared to PuH$_2$.
The H atoms situated between planes sit at almost identical positions,
the additional H atoms in PuH$_3$ sit in the close-packed Pu planes.
Density functional theory (DFT) calculations in the generalized gradient
approximation (GGA) give structural dimensions in good agreement
with experiment.
For the DFT calculations to favor the stacking observed in experiment
requires some treatment of the electron correlation between $5f$ electrons,
exemplified here by the GGA+U method.

The close-packed Pu planes with ABC and AB stacking interspersed with
H atoms provides a framework to explore intermediate stoichiometries.
The experimentally observed hysteresis,
attributed to structural origins,\cite{haschke1987}
can be explained based on the stoichiometric dependence of
which type of stacking has a lower calculated energy (using the GGA+U approach).
The calculated stoichiometry PuH$_x$ at which the preference switches
from ABC stacking to AB stacking, slightly above $x=2.75$,
agrees well with the lower limit of the region in which experiment
observes the two types of crystal structures coexisting.\cite{mulford56}

Overall,
the DFT calculations on the Pu-H system
reveal very small energy differences (relative to room temperature)
for multiple changes to the system:
the AFM solutions lie below the FM solutions by small energies,
changing the arrangements of the H atoms on tetrahedral and octahedral
sites leads to small changes in the energy,
and even changing the stoichiometry itself affects the formation
energy by small amounts.\\

\begin{acknowledgments}

This research was supported by the Los Alamos National Laboratory,
under the auspices of the National Nuclear Security Agency,
by the U.S. Department of Energy under Grant No.
LDRD-DR 20110011 (``Hydrogen Effects in $\delta$-Stabilized Pu Alloys'').
Many thanks go to
Eric Chisolm,
Scott Richmond,
Dan Schwartz,
Dan Sheppard,
Chris Taylor,
Steve Valone,
and
Art Voter
for helpful and encouraging discussions.
The author expresses a deep gratitude to Neil Henson
for assistance with the {\sc andulu} computational facility.
Last, but not least, fond thanks go to Lucia Li\^en and Anna Lan for
spurring alternative approaches to understanding.

\end{acknowledgments}


\begin{thebibliography}{25}%
\makeatletter
\providecommand \@ifxundefined [1]{%
 \@ifx{#1\undefined}
}%
\providecommand \@ifnum [1]{%
 \ifnum #1\expandafter \@firstoftwo
 \else \expandafter \@secondoftwo
 \fi
}%
\providecommand \@ifx [1]{%
 \ifx #1\expandafter \@firstoftwo
 \else \expandafter \@secondoftwo
 \fi
}%
\providecommand \natexlab [1]{#1}%
\providecommand \enquote  [1]{``#1''}%
\providecommand \bibnamefont  [1]{#1}%
\providecommand \bibfnamefont [1]{#1}%
\providecommand \citenamefont [1]{#1}%
\providecommand \href@noop [0]{\@secondoftwo}%
\providecommand \href [0]{\begingroup \@sanitize@url \@href}%
\providecommand \@href[1]{\@@startlink{#1}\@@href}%
\providecommand \@@href[1]{\endgroup#1\@@endlink}%
\providecommand \@sanitize@url [0]{\catcode `\\12\catcode `\$12\catcode
  `\&12\catcode `\#12\catcode `\^12\catcode `\_12\catcode `\%12\relax}%
\providecommand \@@startlink[1]{}%
\providecommand \@@endlink[0]{}%
\providecommand \url  [0]{\begingroup\@sanitize@url \@url }%
\providecommand \@url [1]{\endgroup\@href {#1}{\urlprefix }}%
\providecommand \urlprefix  [0]{URL }%
\providecommand \Eprint [0]{\href }%
\providecommand \doibase [0]{http://dx.doi.org/}%
\providecommand \selectlanguage [0]{\@gobble}%
\providecommand \bibinfo  [0]{\@secondoftwo}%
\providecommand \bibfield  [0]{\@secondoftwo}%
\providecommand \translation [1]{[#1]}%
\providecommand \BibitemOpen [0]{}%
\providecommand \bibitemStop [0]{}%
\providecommand \bibitemNoStop [0]{.\EOS\space}%
\providecommand \EOS [0]{\spacefactor3000\relax}%
\providecommand \BibitemShut  [1]{\csname bibitem#1\endcsname}%
\let\auto@bib@innerbib\@empty
\bibitem [{\citenamefont {Martz}\ and\ \citenamefont
  {Schwartz}(2003)}]{martz03}%
  \BibitemOpen
  \bibfield  {author} {\bibinfo {author} {\bibfnamefont {J.}~\bibnamefont
  {Martz}}\ and\ \bibinfo {author} {\bibfnamefont {A.}~\bibnamefont
  {Schwartz}},\ }\href@noop {} {\bibfield  {journal} {\bibinfo  {journal} {JOM
  Journal of the Minerals, Metals and Materials Society}\ }\textbf {\bibinfo
  {volume} {55}},\ \bibinfo {pages} {19} (\bibinfo {year} {2003})}\BibitemShut
  {NoStop}%
\bibitem [{\citenamefont {Richmond}\ \emph {et~al.}(2010)\citenamefont
  {Richmond}, \citenamefont {Bridgewater}, \citenamefont {Ward},\ and\
  \citenamefont {Allen}}]{richmond2010}%
  \BibitemOpen
  \bibfield  {author} {\bibinfo {author} {\bibfnamefont {S.}~\bibnamefont
  {Richmond}}, \bibinfo {author} {\bibfnamefont {J.~S.}\ \bibnamefont
  {Bridgewater}}, \bibinfo {author} {\bibfnamefont {J.~W.}\ \bibnamefont
  {Ward}}, \ and\ \bibinfo {author} {\bibfnamefont {T.~H.}\ \bibnamefont
  {Allen}},\ }\href@noop {} {\bibfield  {journal} {\bibinfo  {journal} {IOP
  Conference Series: Materials Science and Engineering}\ }\textbf {\bibinfo
  {volume} {9}},\ \bibinfo {pages} {012036} (\bibinfo {year}
  {2010})}\BibitemShut {NoStop}%
\bibitem [{\citenamefont {Haschke}(1991)}]{haschke91}%
  \BibitemOpen
  \bibfield  {author} {\bibinfo {author} {\bibfnamefont {J.~M.}\ \bibnamefont
  {Haschke}},\ }\enquote {\bibinfo {title} {Synthesis of lanthanide and
  actinide compounds},}\ \ (\bibinfo  {publisher} {Kluwer},\ \bibinfo {address}
  {Dordrecht},\ \bibinfo {year} {1991})\ p.~\bibinfo {pages} {1}\BibitemShut
  {NoStop}%
\bibitem [{\citenamefont {Haschke}\ and\ \citenamefont
  {Allen}(2001)}]{haschke2001}%
  \BibitemOpen
  \bibfield  {author} {\bibinfo {author} {\bibfnamefont {J.~M.}\ \bibnamefont
  {Haschke}}\ and\ \bibinfo {author} {\bibfnamefont {T.~H.}\ \bibnamefont
  {Allen}},\ }\href@noop {} {\bibfield  {journal} {\bibinfo  {journal} {Journal
  of Alloys and Compounds}\ }\textbf {\bibinfo {volume} {320}},\ \bibinfo
  {pages} {58 } (\bibinfo {year} {2001})}\BibitemShut {NoStop}%
\bibitem [{\citenamefont {Eriksson}\ \emph {et~al.}(1991)\citenamefont
  {Eriksson}, \citenamefont {Hao}, \citenamefont {Cooper}, \citenamefont
  {Fernando}, \citenamefont {Cox}, \citenamefont {Ward},\ and\ \citenamefont
  {Boring}}]{eriksson91}%
  \BibitemOpen
  \bibfield  {author} {\bibinfo {author} {\bibfnamefont {O.}~\bibnamefont
  {Eriksson}}, \bibinfo {author} {\bibfnamefont {Y.~G.}\ \bibnamefont {Hao}},
  \bibinfo {author} {\bibfnamefont {B.~R.}\ \bibnamefont {Cooper}}, \bibinfo
  {author} {\bibfnamefont {G.~W.}\ \bibnamefont {Fernando}}, \bibinfo {author}
  {\bibfnamefont {L.~E.}\ \bibnamefont {Cox}}, \bibinfo {author} {\bibfnamefont
  {J.~W.}\ \bibnamefont {Ward}}, \ and\ \bibinfo {author} {\bibfnamefont
  {A.~M.}\ \bibnamefont {Boring}},\ }\href@noop {} {\bibfield  {journal}
  {\bibinfo  {journal} {Phys. Rev. B}\ }\textbf {\bibinfo {volume} {43}},\
  \bibinfo {pages} {4590} (\bibinfo {year} {1991})}\BibitemShut {NoStop}%
\bibitem [{\citenamefont {Ai}\ \emph {et~al.}(2012)\citenamefont {Ai},
  \citenamefont {Liu}, \citenamefont {Gao},\ and\ \citenamefont {Ao}}]{Ai2012}%
  \BibitemOpen
  \bibfield  {author} {\bibinfo {author} {\bibfnamefont {J.}~\bibnamefont
  {Ai}}, \bibinfo {author} {\bibfnamefont {T.}~\bibnamefont {Liu}}, \bibinfo
  {author} {\bibfnamefont {T.}~\bibnamefont {Gao}}, \ and\ \bibinfo {author}
  {\bibfnamefont {B.}~\bibnamefont {Ao}},\ }\href@noop {} {\bibfield  {journal}
  {\bibinfo  {journal} {Computational Materials Science}\ }\textbf {\bibinfo
  {volume} {51}},\ \bibinfo {pages} {127 } (\bibinfo {year}
  {2012})}\BibitemShut {NoStop}%
\bibitem [{\citenamefont {Ao}\ \emph {et~al.}(2012)\citenamefont {Ao},
  \citenamefont {Wang}, \citenamefont {Shi}, \citenamefont {Chen},
  \citenamefont {Ye}, \citenamefont {Lai},\ and\ \citenamefont
  {Gao}}]{Ao2012jnm}%
  \BibitemOpen
  \bibfield  {author} {\bibinfo {author} {\bibfnamefont {B.}~\bibnamefont
  {Ao}}, \bibinfo {author} {\bibfnamefont {X.}~\bibnamefont {Wang}}, \bibinfo
  {author} {\bibfnamefont {P.}~\bibnamefont {Shi}}, \bibinfo {author}
  {\bibfnamefont {P.}~\bibnamefont {Chen}}, \bibinfo {author} {\bibfnamefont
  {X.}~\bibnamefont {Ye}}, \bibinfo {author} {\bibfnamefont {X.}~\bibnamefont
  {Lai}}, \ and\ \bibinfo {author} {\bibfnamefont {T.}~\bibnamefont {Gao}},\
  }\href@noop {} {\bibfield  {journal} {\bibinfo  {journal} {Journal of Nuclear
  Materials}\ }\textbf {\bibinfo {volume} {424}},\ \bibinfo {pages} {183 }
  (\bibinfo {year} {2012})}\BibitemShut {NoStop}%
\bibitem [{\citenamefont {Bing-Yun}\ \emph {et~al.}(2012)\citenamefont
  {Bing-Yun}, \citenamefont {Juan-Juan}, \citenamefont {Tao}, \citenamefont
  {Xiao-Lin}, \citenamefont {Peng}, \citenamefont {Pi-Heng},\ and\
  \citenamefont {Xiao-Qiu}}]{Ao2012cpl}%
  \BibitemOpen
  \bibfield  {author} {\bibinfo {author} {\bibfnamefont {A.}~\bibnamefont
  {Bing-Yun}}, \bibinfo {author} {\bibfnamefont {A.}~\bibnamefont {Juan-Juan}},
  \bibinfo {author} {\bibfnamefont {G.}~\bibnamefont {Tao}}, \bibinfo {author}
  {\bibfnamefont {W.}~\bibnamefont {Xiao-Lin}}, \bibinfo {author}
  {\bibfnamefont {S.}~\bibnamefont {Peng}}, \bibinfo {author} {\bibfnamefont
  {C.}~\bibnamefont {Pi-Heng}}, \ and\ \bibinfo {author} {\bibfnamefont
  {Y.}~\bibnamefont {Xiao-Qiu}},\ }\href@noop {} {\bibfield  {journal}
  {\bibinfo  {journal} {Chinese Physics Letters}\ }\textbf {\bibinfo {volume}
  {29}},\ \bibinfo {pages} {017102} (\bibinfo {year} {2012})}\BibitemShut
  {NoStop}%
\bibitem [{\citenamefont {Mulford}\ and\ \citenamefont
  {Sturdy}(1955)}]{mulford55}%
  \BibitemOpen
  \bibfield  {author} {\bibinfo {author} {\bibfnamefont {R.~N.~R.}\
  \bibnamefont {Mulford}}\ and\ \bibinfo {author} {\bibfnamefont {G.~E.}\
  \bibnamefont {Sturdy}},\ }\href@noop {} {\bibfield  {journal} {\bibinfo
  {journal} {Journal of the American Chemical Society}\ }\textbf {\bibinfo
  {volume} {77}},\ \bibinfo {pages} {3449} (\bibinfo {year}
  {1955})}\BibitemShut {NoStop}%
\bibitem [{\citenamefont {Mulford}\ and\ \citenamefont
  {Sturdy}(1956)}]{mulford56}%
  \BibitemOpen
  \bibfield  {author} {\bibinfo {author} {\bibfnamefont {R.~N.~R.}\
  \bibnamefont {Mulford}}\ and\ \bibinfo {author} {\bibfnamefont {G.~E.}\
  \bibnamefont {Sturdy}},\ }\href@noop {} {\bibfield  {journal} {\bibinfo
  {journal} {Journal of the American Chemical Society}\ }\textbf {\bibinfo
  {volume} {78}},\ \bibinfo {pages} {3897} (\bibinfo {year}
  {1956})}\BibitemShut {NoStop}%
\bibitem [{\citenamefont {Aldred}\ \emph {et~al.}(1979)\citenamefont {Aldred},
  \citenamefont {Cinadar}, \citenamefont {Lam},\ and\ \citenamefont
  {Weber}}]{aldred1979}%
  \BibitemOpen
  \bibfield  {author} {\bibinfo {author} {\bibfnamefont {A.~T.}\ \bibnamefont
  {Aldred}}, \bibinfo {author} {\bibfnamefont {G.}~\bibnamefont {Cinadar}},
  \bibinfo {author} {\bibfnamefont {D.~J.}\ \bibnamefont {Lam}}, \ and\
  \bibinfo {author} {\bibfnamefont {L.~W.}\ \bibnamefont {Weber}},\ }\href@noop
  {} {\bibfield  {journal} {\bibinfo  {journal} {Phys. Rev. B}\ }\textbf
  {\bibinfo {volume} {19}},\ \bibinfo {pages} {300} (\bibinfo {year}
  {1979})}\BibitemShut {NoStop}%
\bibitem [{\citenamefont {Ward}(1983)}]{ward1983}%
  \BibitemOpen
  \bibfield  {author} {\bibinfo {author} {\bibfnamefont {J.~W.}\ \bibnamefont
  {Ward}},\ }\href@noop {} {\bibfield  {journal} {\bibinfo  {journal} {Journal
  of the Less Common Metals}\ }\textbf {\bibinfo {volume} {93}},\ \bibinfo
  {pages} {279 } (\bibinfo {year} {1983})}\BibitemShut {NoStop}%
\bibitem [{\citenamefont {Willis}\ \emph {et~al.}(1985)\citenamefont {Willis},
  \citenamefont {Ward}, \citenamefont {Smith}, \citenamefont {Kosiewicz},
  \citenamefont {Haschke},\ and\ \citenamefont {III}}]{willis1985}%
  \BibitemOpen
  \bibfield  {author} {\bibinfo {author} {\bibfnamefont {J.}~\bibnamefont
  {Willis}}, \bibinfo {author} {\bibfnamefont {J.}~\bibnamefont {Ward}},
  \bibinfo {author} {\bibfnamefont {J.}~\bibnamefont {Smith}}, \bibinfo
  {author} {\bibfnamefont {S.}~\bibnamefont {Kosiewicz}}, \bibinfo {author}
  {\bibfnamefont {J.}~\bibnamefont {Haschke}}, \ and\ \bibinfo {author}
  {\bibfnamefont {A.~H.}\ \bibnamefont {III}},\ }\href@noop {} {\bibfield
  {journal} {\bibinfo  {journal} {Physica B+C}\ }\textbf {\bibinfo {volume}
  {130}},\ \bibinfo {pages} {527 } (\bibinfo {year} {1985})}\BibitemShut
  {NoStop}%
\bibitem [{\citenamefont {Haschke}\ \emph {et~al.}(1987)\citenamefont
  {Haschke}, \citenamefont {III},\ and\ \citenamefont {Lucas}}]{haschke1987}%
  \BibitemOpen
  \bibfield  {author} {\bibinfo {author} {\bibfnamefont {J.~M.}\ \bibnamefont
  {Haschke}}, \bibinfo {author} {\bibfnamefont {A.~E.~H.}\ \bibnamefont {III}},
  \ and\ \bibinfo {author} {\bibfnamefont {R.~L.}\ \bibnamefont {Lucas}},\
  }\href@noop {} {\bibfield  {journal} {\bibinfo  {journal} {Journal of the
  Less Common Metals}\ }\textbf {\bibinfo {volume} {133}},\ \bibinfo {pages}
  {155 } (\bibinfo {year} {1987})}\BibitemShut {NoStop}%
\bibitem [{\citenamefont {McGillivray}\ \emph {et~al.}(2003)\citenamefont
  {McGillivray}, \citenamefont {Findlay}, \citenamefont {Harker},\ and\
  \citenamefont {Trask}}]{mcgillivray2003}%
  \BibitemOpen
  \bibfield  {author} {\bibinfo {author} {\bibfnamefont {G.~W.}\ \bibnamefont
  {McGillivray}}, \bibinfo {author} {\bibfnamefont {I.~M.}\ \bibnamefont
  {Findlay}}, \bibinfo {author} {\bibfnamefont {R.~M.}\ \bibnamefont {Harker}},
  \ and\ \bibinfo {author} {\bibfnamefont {I.~D.}\ \bibnamefont {Trask}},\
  }\href@noop {} {\bibfield  {journal} {\bibinfo  {journal} {AIP Conference
  Proceedings}\ }\textbf {\bibinfo {volume} {673}},\ \bibinfo {pages} {167}
  (\bibinfo {year} {2003})}\BibitemShut {NoStop}%
\bibitem [{\citenamefont {Crocker}(1971)}]{crocker1971}%
  \BibitemOpen
  \bibfield  {author} {\bibinfo {author} {\bibfnamefont {A.}~\bibnamefont
  {Crocker}},\ }\href@noop {} {\bibfield  {journal} {\bibinfo  {journal}
  {Journal of Nuclear Materials}\ }\textbf {\bibinfo {volume} {41}},\ \bibinfo
  {pages} {167 } (\bibinfo {year} {1971})}\BibitemShut {NoStop}%
\bibitem [{\citenamefont {Kresse}\ and\ \citenamefont
  {Furthmuller}(1996)}]{kresse96}%
  \BibitemOpen
  \bibfield  {author} {\bibinfo {author} {\bibfnamefont {G.}~\bibnamefont
  {Kresse}}\ and\ \bibinfo {author} {\bibfnamefont {J.}~\bibnamefont
  {Furthmuller}},\ }\href@noop {} {\bibfield  {journal} {\bibinfo  {journal}
  {Phys. Rev. B}\ }\textbf {\bibinfo {volume} {54}},\ \bibinfo {pages} {11169}
  (\bibinfo {year} {1996})}\BibitemShut {NoStop}%
\bibitem [{\citenamefont {Kresse}\ and\ \citenamefont
  {Joubert}(1999)}]{kresse99}%
  \BibitemOpen
  \bibfield  {author} {\bibinfo {author} {\bibfnamefont {G.}~\bibnamefont
  {Kresse}}\ and\ \bibinfo {author} {\bibfnamefont {D.}~\bibnamefont
  {Joubert}},\ }\href@noop {} {\bibfield  {journal} {\bibinfo  {journal} {Phys.
  Rev. B}\ }\textbf {\bibinfo {volume} {59}},\ \bibinfo {pages} {1758}
  (\bibinfo {year} {1999})}\BibitemShut {NoStop}%
\bibitem [{\citenamefont {Perdew}\ \emph {et~al.}(1996)\citenamefont {Perdew},
  \citenamefont {Burke},\ and\ \citenamefont {Ernzerhof}}]{PBE96}%
  \BibitemOpen
  \bibfield  {author} {\bibinfo {author} {\bibfnamefont {J.~P.}\ \bibnamefont
  {Perdew}}, \bibinfo {author} {\bibfnamefont {K.}~\bibnamefont {Burke}}, \
  and\ \bibinfo {author} {\bibfnamefont {M.}~\bibnamefont {Ernzerhof}},\
  }\href@noop {} {\bibfield  {journal} {\bibinfo  {journal} {Phys.\ Rev.\
  Lett.}\ }\textbf {\bibinfo {volume} {77}},\ \bibinfo {pages} {3865} (\bibinfo
  {year} {1996})}\BibitemShut {NoStop}%
\bibitem [{\citenamefont {Bl{\"o}chl}(1994)}]{blochl94a}%
  \BibitemOpen
  \bibfield  {author} {\bibinfo {author} {\bibfnamefont {P.~E.}\ \bibnamefont
  {Bl{\"o}chl}},\ }\href@noop {} {\bibfield  {journal} {\bibinfo  {journal}
  {Phys. Rev. B}\ }\textbf {\bibinfo {volume} {50}},\ \bibinfo {pages} {17953}
  (\bibinfo {year} {1994})}\BibitemShut {NoStop}%
\bibitem [{\citenamefont {Bl{\"o}chl}\ \emph {et~al.}(1994)\citenamefont
  {Bl{\"o}chl}, \citenamefont {Jepsen},\ and\ \citenamefont
  {Andersen}}]{blochl94}%
  \BibitemOpen
  \bibfield  {author} {\bibinfo {author} {\bibfnamefont {P.~E.}\ \bibnamefont
  {Bl{\"o}chl}}, \bibinfo {author} {\bibfnamefont {O.}~\bibnamefont {Jepsen}},
  \ and\ \bibinfo {author} {\bibfnamefont {O.~K.}\ \bibnamefont {Andersen}},\
  }\href@noop {} {\bibfield  {journal} {\bibinfo  {journal} {Phys. Rev. B}\
  }\textbf {\bibinfo {volume} {49}},\ \bibinfo {pages} {16223} (\bibinfo {year}
  {1994})}\BibitemShut {NoStop}%
\bibitem [{\citenamefont {Dudarev}\ \emph {et~al.}(1998)\citenamefont
  {Dudarev}, \citenamefont {Botton}, \citenamefont {Savrasov}, \citenamefont
  {Humphreys},\ and\ \citenamefont {Sutton}}]{dudarev98}%
  \BibitemOpen
  \bibfield  {author} {\bibinfo {author} {\bibfnamefont {S.~L.}\ \bibnamefont
  {Dudarev}}, \bibinfo {author} {\bibfnamefont {G.~A.}\ \bibnamefont {Botton}},
  \bibinfo {author} {\bibfnamefont {S.~Y.}\ \bibnamefont {Savrasov}}, \bibinfo
  {author} {\bibfnamefont {C.~J.}\ \bibnamefont {Humphreys}}, \ and\ \bibinfo
  {author} {\bibfnamefont {A.~P.}\ \bibnamefont {Sutton}},\ }\href@noop {}
  {\bibfield  {journal} {\bibinfo  {journal} {Phys. Rev. B}\ }\textbf {\bibinfo
  {volume} {57}},\ \bibinfo {pages} {1505} (\bibinfo {year}
  {1998})}\BibitemShut {NoStop}%
\bibitem [{\citenamefont {Bartscher}\ \emph {et~al.}(1984)\citenamefont
  {Bartscher}, \citenamefont {Boeuf}, \citenamefont {Caciuffo}, \citenamefont
  {Fournier}, \citenamefont {Haschke}, \citenamefont {Manes}, \citenamefont
  {Rebizant}, \citenamefont {Rustichelli},\ and\ \citenamefont
  {Ward}}]{bartscher1984}%
  \BibitemOpen
  \bibfield  {author} {\bibinfo {author} {\bibfnamefont {W.}~\bibnamefont
  {Bartscher}}, \bibinfo {author} {\bibfnamefont {A.}~\bibnamefont {Boeuf}},
  \bibinfo {author} {\bibfnamefont {R.}~\bibnamefont {Caciuffo}}, \bibinfo
  {author} {\bibfnamefont {J.}~\bibnamefont {Fournier}}, \bibinfo {author}
  {\bibfnamefont {J.}~\bibnamefont {Haschke}}, \bibinfo {author} {\bibfnamefont
  {L.}~\bibnamefont {Manes}}, \bibinfo {author} {\bibfnamefont
  {J.}~\bibnamefont {Rebizant}}, \bibinfo {author} {\bibfnamefont
  {F.}~\bibnamefont {Rustichelli}}, \ and\ \bibinfo {author} {\bibfnamefont
  {J.}~\bibnamefont {Ward}},\ }\href@noop {} {\bibfield  {journal} {\bibinfo
  {journal} {Solid State Communications}\ }\textbf {\bibinfo {volume} {52}},\
  \bibinfo {pages} {619 } (\bibinfo {year} {1984})}\BibitemShut {NoStop}%
\bibitem [{\citenamefont {Bartscher}\ \emph {et~al.}(1985)\citenamefont
  {Bartscher}, \citenamefont {Boeuf}, \citenamefont {Caciuffo}, \citenamefont
  {Fournier}, \citenamefont {Kuhs}, \citenamefont {Rebizant},\ and\
  \citenamefont {Rustichelli}}]{bartscher1985}%
  \BibitemOpen
  \bibfield  {author} {\bibinfo {author} {\bibfnamefont {W.}~\bibnamefont
  {Bartscher}}, \bibinfo {author} {\bibfnamefont {A.}~\bibnamefont {Boeuf}},
  \bibinfo {author} {\bibfnamefont {R.}~\bibnamefont {Caciuffo}}, \bibinfo
  {author} {\bibfnamefont {J.}~\bibnamefont {Fournier}}, \bibinfo {author}
  {\bibfnamefont {W.}~\bibnamefont {Kuhs}}, \bibinfo {author} {\bibfnamefont
  {J.}~\bibnamefont {Rebizant}}, \ and\ \bibinfo {author} {\bibfnamefont
  {F.}~\bibnamefont {Rustichelli}},\ }\href@noop {} {\bibfield  {journal}
  {\bibinfo  {journal} {Solid State Communications}\ }\textbf {\bibinfo
  {volume} {53}},\ \bibinfo {pages} {423 } (\bibinfo {year}
  {1985})}\BibitemShut {NoStop}%
\bibitem [{\citenamefont {Sheppard}\ \emph {et~al.}(2012)\citenamefont
  {Sheppard}, \citenamefont {Xiao}, \citenamefont {Chemelewski}, \citenamefont
  {Johnson},\ and\ \citenamefont {Henkelman}}]{Sheppard2012}%
  \BibitemOpen
  \bibfield  {author} {\bibinfo {author} {\bibfnamefont {D.}~\bibnamefont
  {Sheppard}}, \bibinfo {author} {\bibfnamefont {P.}~\bibnamefont {Xiao}},
  \bibinfo {author} {\bibfnamefont {W.}~\bibnamefont {Chemelewski}}, \bibinfo
  {author} {\bibfnamefont {D.~D.}\ \bibnamefont {Johnson}}, \ and\ \bibinfo
  {author} {\bibfnamefont {G.}~\bibnamefont {Henkelman}},\ }\href@noop {}
  {\bibfield  {journal} {\bibinfo  {journal} {The Journal of Chemical Physics}\
  }\textbf {\bibinfo {volume} {136}},\ \bibinfo {pages} {074103} (\bibinfo
  {year} {2012})}\BibitemShut {NoStop}%
\end{thebibliography}
\end{document}